\documentclass{emulateapj} 
\usepackage{apjfonts}

\def\msun{M$_\odot$}

\newcommand{\mum}{\ifmmode{\rm \mu m}\else{$\mu$m}\fi}

\shorttitle{Ne and S Abundances of PNe in the Magellanic Clouds}
\shortauthors{Bernard-Salas et al.}

\begin{document}

\title{Neon and Sulfur Abundances
  of Planetary Nebulae in the  Magellanic Clouds}

\author{J.~Bernard-Salas\altaffilmark{1},
S.~R.~Pottasch\altaffilmark{2},
S.~Gutenkunst,\altaffilmark{1},
P.~W.~Morris\altaffilmark{3},
J.~R.~Houck\altaffilmark{1}
}


\altaffiltext{1}{Center for Radiophysics and Space Research, Cornell
University, 222 Space Sciences Building, Ithaca, NY 14853-6801, USA.}
\altaffiltext{2}{Kapteyn Astronomical Institute, 9700 AV, Groningen,
The Netherlands.}
\altaffiltext{3}{NASA Herschel Science Center, IPAC/Caltech, MS
100-22, Pasadena, CA 91125.}

\begin{abstract}
  
  The chemical abundances of neon and sulfur for 25 planetary nebulae
  (PNe) in the Magellanic Clouds are presented. These abundances have
  been derived using mainly infrared data from the {\em Spitzer Space
    Telescope}.  The implications for the chemical evolution of these
  elements are discussed. A comparison with similarly obtained
  abundances of Galactic PNe and H\,II regions and Magellanic Clouds
  H\,II regions is also given.  The average neon abundances are
  6.0$\times$10$^{-5}$ and 2.7$\times$10$^{-5}$ for the PNe in the
  Large and Small Magellanic Clouds respectively.  These are $\sim$1/3
  and 1/6 of the average abundances of Galactic planetary nebulae to
  which we compare.  The average sulfur abundances for the LMC and SMC
  are respectively 2.7$\times$10$^{-6}$ and 1.0$\times$10$^{-6}$.  The
  Ne/S ratio (23.5) is on average higher than the ratio found in
  Galactic PNe (16) but the range of values in both data sets is
  similar for most of the objects.  The neon abundances found in PNe
  and H\,II regions agree with each other. It is possible that a few
  (3-4) of the PNe in the sample have experienced some neon
  enrichment, but for two of these objects the high Ne/S ratio can be
  explained by their very low sulfur abundances.  The neon and sulfur
  abundances derived in this paper are also compared to previously
  published abundances using optical data and photo-ionization models.

\end{abstract}


\section{Introduction}

Stars of low- and intermediate-mass ($\sim$1-8~\msun) become planetary
nebulae after they evolve off the Asymptotic Giant Branch (AGB)
\citep{ibe}. In the PN phase the hot central star ionizes the
previously ejected material which then emits copious amounts of
emission lines of different ions. These emission lines are ideal to
study the chemical composition of the gas. The abundances of elements
such as carbon, nitrogen and oxygen can be used to give information on
the nucleosynthesis history of the progenitor star. Other elements,
such as neon, sulfur and argon are not supposed to be altered in the
course of evolution of low- and intermediate-mass stars and are
therefore indicators of the chemical composition at the epoch of
formation \citep{mar}.

It is for these reasons that PNe have been the subject of many
spectroscopic studies over the years.  Due to limitations in the
observations, the bulk of this spectroscopic work has been focused on
analysing PNe in the Milky Way (MW). However, observations of PNe
outside the Galaxy are very important because one can probe different
metallicity regions and, unlike Galactic PNe, the distance is known
which allows one to relate the abundance to the central star
luminosity. During the last years several papers \citep[e.g.][ and
references therein]{mag01,mag03,cor05} have been devoted to
identifying PNe outside the Galaxy. As a consequence, the number of
PNe known in the Local Group keeps increasing.  The Large and Small
Magellanic Clouds (hereafter LMC and SMC respectively) are ideal
candidates to obtain spectroscopic observations of PNe.

\citet{all81} and \citet{all83} published optical spectroscopic data
for 6 PNe in the LMC and 7 in SMC respectively. In a follow-up paper
\citet{all87} presented ultraviolet data from the {\em IUE} satellite
of 12 PNe in the Magellanic Clouds (MC) and derived abundances for
several elements.  Ultraviolet data are essential in order to derive
abundances of elements such as carbon and nitrogen. Optical
spectroscopy and abundances of 71 MC PNe were presented by
\citet{mon88}. In the early 90's \citet{mea91a, mea91b} obtained
optical spectroscopy of over a hundred PNe in the MC.  \citet{mor}
presented FLAIR spectroscopy of 97 PNe in the LMC which included
fluxes of the \ion{[O}{3]}, \ion{[S}{2]}, \ion{[N}{2]}, and
\ion{He}{2} lines.  Using optical and IUE data, \citet{pen97} studied
a sample of MC PNe with WR nuclei and found that the distribution of
spectral type was different from those of Galactic WR-PNe.
\citet{sta02,sta03} have characterised optically a large number of PNe
in the Magellanic Clouds using {\em HST} observations. And more
recently \citet{lei06} have derived the abundances of several elements
for a large sample of PNe in the MC using optical data.

Despite their importance, infrared spectroscopic studies of PNe in the
MC are scarce in the literature. This is mainly because full
integrated spectra in the infrared can only be achieved from space.
IRAS detected several PNe in the MC \citep{zij94}, mainly at 12 and
25~$\mu$m, but some of the identifications are dubious. The SWS
spectrograph \citep{deg} on board {\em ISO} did an excellent job
studying nearby PNe, but it did not have enough sensitivity to allow
the study of PNe outside the Galaxy. The {\em Spitzer Space Telescope}
\citep{wer} with its increased sensitivity enables us to observe PNe
outside the Milky Way \citep[][]{ber06,ber04}. The importance of using
infrared lines when deriving abundances has been highlighted by
\citet{rub} and we summarize these reasons here. Infrared lines are
little affected by extinction as opposed to optical or UV lines.
Uncertainties in the electron temperature or fluctuations in the
temperature within the nebula are not important when using infrared
lines because they originate from levels very close to the ground
level.  Additionally many ions emit in the infrared, and therefore the
use of ionization correction factors (ICFs) can be greatly reduced by
including infrared observations. This is especially true for neon,
sulfur and argon.  Finally, and while not discussed in this paper,
emission of dust can be studied in this part of the electromagnetic
spectrum.

In this paper we present {\it Spitzer} high-resolution spectroscopic
observations of 25 PNe in the MC (18 in the LMC and 7 in the SMC).
This paper focuses on the emission of fine-structure lines and their
use in the abundance determination.  The measured lines are used to
derive abundances for sulfur and neon.  These abundances are mainly
compared to Galactic PNe abundances from \cite{pot06}, as well as
Galactic, MC, M33 and M83 H\,II regions abundances from \citet{leti},
\citet{ver}, and \citet{rub07}. All of these abundances have been
derived using infrared data and in a similar way to that presented in
this paper.  Dust features present in the spectra such as Polycyclic
Aromatic Hydrocarbons (PAHs) and silicates will be discussed in a
future paper.

\section{Observations and Data Reduction}
\label{obs_s}

\begin{figure}
  \begin{center}
  \includegraphics[width=8cm]{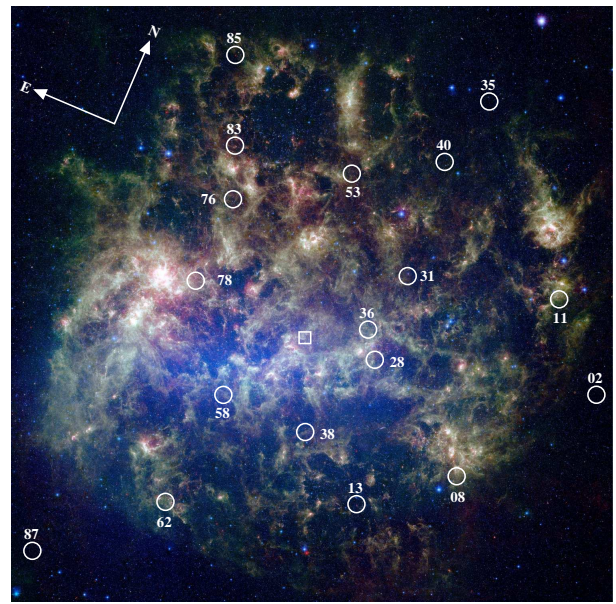}
  \end{center}
  \caption{IRAC four-band mosaic of the LMC (SAGE) with the positions
    of the PNe overlaid on it (circles). The dynamical center of the
    LMC as given by \cite{kim} is indicated by the square close to the
    center of the figure.
    \label{lmcirac_f}}
\end{figure}

The observations were made using the Infrared Spectrograph (IRS)
\citep{hou} on board the {\it Spitzer Space Telescope} and resulted in
high- and low-resolution spectra of 25 PNe.  These observations were
part of the GTO program (ID 103) and were taken between March and
November 2005.  The object name and AORkey numbers for each
observation are given in the first and second columns of
Table~\ref{tene_t}. The nomenclature given by \cite{san} is adopted
through the paper. In addition, the analysis includes data on
SMP~LMC~31, and abundances on SMP~LMC~83 derived by \citet{ber04}.
These data were taken during In Orbit Check-Out (IOC). There are two
high-resolution modules in the IRS, named Short-High and Long-High (SH
and LH respectively). Together, they cover the wavelength region
between 10 and 37~$\mu$m at a resolution of 600. The reader should
refer to the paper by \citet{hou} for more information on the IRS
instrument. We used coordinates given by \citet{sta02, sta03} and
\citet{lei97}, and performed blue Peak-Up acquisition on a nearby star
to obtain accurate pointing (0.4\arcsec).  Figure~\ref{lmcirac_f}
shows the position of the LMC PNe on an IRAC image from the SAGE team
\citep[][]{mei}.

\begin{deluxetable}{c c c c c c}
  \tablewidth{0pc}
  \tablecaption{Adopted parameters for abundance determination.\label{tene_t}}

  \tablehead{\colhead{Object} & \colhead{AORkey} &
  \colhead{Log(F$_{H\beta}$)\tablenotemark{a,b}} &
  \colhead{C$_{H\beta}$\tablenotemark{a}} & \colhead{T$_e$ (K)\tablenotemark{a}} & \colhead{N$_e$
  (cm$^{-3}$)\tablenotemark{a}}}

  \startdata    

  SMP LMC~02  &  4946944  &  -13.18 &  0.06  &   11600  &   3000\tablenotemark{c} \\
  SMP LMC~08  &  15902464 &  -13.74 &  0.23  &   11000  &   5500  \\
  SMP LMC~11  &  4947712  &  -13.94 &  0.31  &   25000  &   6200  \\ 
  SMP LMC~13  &  4947968  &  -12.88 &  0.09  &   13600  &   3800  \\
  SMP LMC~28  &  4948224  &  -13.57 &  0.32  &   10000  &   2000  \\
  SMP LMC~31  &  7459584  &  -12.92 &  0.54  &   12800  &   6800  \\
  SMP LMC~35  &  4948736  &  -12.81 &  0.04  &   13300  &   1600  \\
  SMP LMC~36  &  4949248  &  -12.72 &  0.41  &   15000  &   3000\tablenotemark{c} \\
  SMP LMC~38  &  12633600 &  -12.62 &  0.21  &   13000  &   9800  \\
  SMP LMC~40  &  4949504  &  -13.25 &  0.20  &   13900  &   1100  \\
  SMP LMC~53  &  15902720 &  -12.62 &  0.13  &   13700  &   4000  \\
  SMP LMC~58  &  4950784  &  -12.54 &  0.11  &   12100  &  20000  \\
  SMP LMC~61  &  12633856 &  -12.48 &  0.22  &   10800  &  26000  \\ 
  SMP LMC~62  &  4951040  &  -12.30 &  0.07  &   15800  &   4400  \\
  SMP LMC~76  &  4951296  &  -12.54 &  0.34  &   11600  &  13600  \\
  SMP LMC~78  &  15902208 &  -12.60 &  0.21  &   14200  &   4300  \\
  SMP LMC~85  &  4952320  &  -12.42 &  0.26  &   10500  &  31400  \\
  SMP LMC~87  &  4952576  &  -12.91 &  0.25  &   19200  &   1900  \\
              &		  & 	    &        &          &         \\
  SMP SMC~01  &  4953088  &  -12.85 &  0.287 &   11000  &   9600  \\
  SMP SMC~03  &  4953600  &  -13.13 &  0.000 &   13800  &   5600  \\
  SMP SMC~06  &  4954112  &  -12.80 &  0.385 &   15300  &  14700  \\
  SMP SMC~11  &  15902976 &  -12.87 &  0.82  &   17600  &   1100  \\
  SMP SMC~22  &  4954624  &  -12.94 &  0.165 &   18800  &   2500  \\
  SMP SMC~24  &  15901952 &  -12.66 &  0.047 &   12700  &   1300  \\ 
  SMP SMC~28  &  4955136  &  -13.18 &  0.200 &   20300  &   8800  \\  

  \enddata

  \tablenotetext{a}{Values taken from the following references (see
    \S~3.2): \citet{lei06, mea88, mea91a, mea91b, sha06, sta02,
      sta05, vil03, vil04, woo87}.}
  \tablenotetext{b}{Flux in units of erg~s$^{-1}$~cm$^{-2}$.}
  \tablenotetext{c}{Assumed electron density.}

\end{deluxetable}

The data were processed through a copy of the S13.2 version of the
{\it Spitzer} Science Center's pipeline which is maintained at Cornell
and using a script version of {\it Smart} \citep{hig}.  The reduction
started from the {\it droop} images. These are equivalent to the most
commonly used {\it bcd} data and only lack the flatfield and
stray-cross-light removal (which is only important for bright
sources).  Rogue pixels which are especially notorious in the LH
module were removed using the {\it irsclean}\footnote{This tool is
  available from the SSC web site: http://ssc.spitzer.caltech.edu}
tool.  The rogue pixels were first flagged using a campaign mask and
then removed. If different cycles (repetitions) were present for a
given observation these were combined using the mean to improve the
S/N. The 2D-images were extracted using full aperture extraction. The
calibration was performed by dividing the resultant spectrum by that
of the calibration star $\xi$dra (extracted in the same way as the
target) and multiplying by its template \citep[][ and Sloan et al. in
prep]{coh}. Finally, glitches which were not present in both nod
positions or in the overlapping region between orders were removed
manually.

\begin{figure*}
  \begin{center}
    \includegraphics[width=15cm]{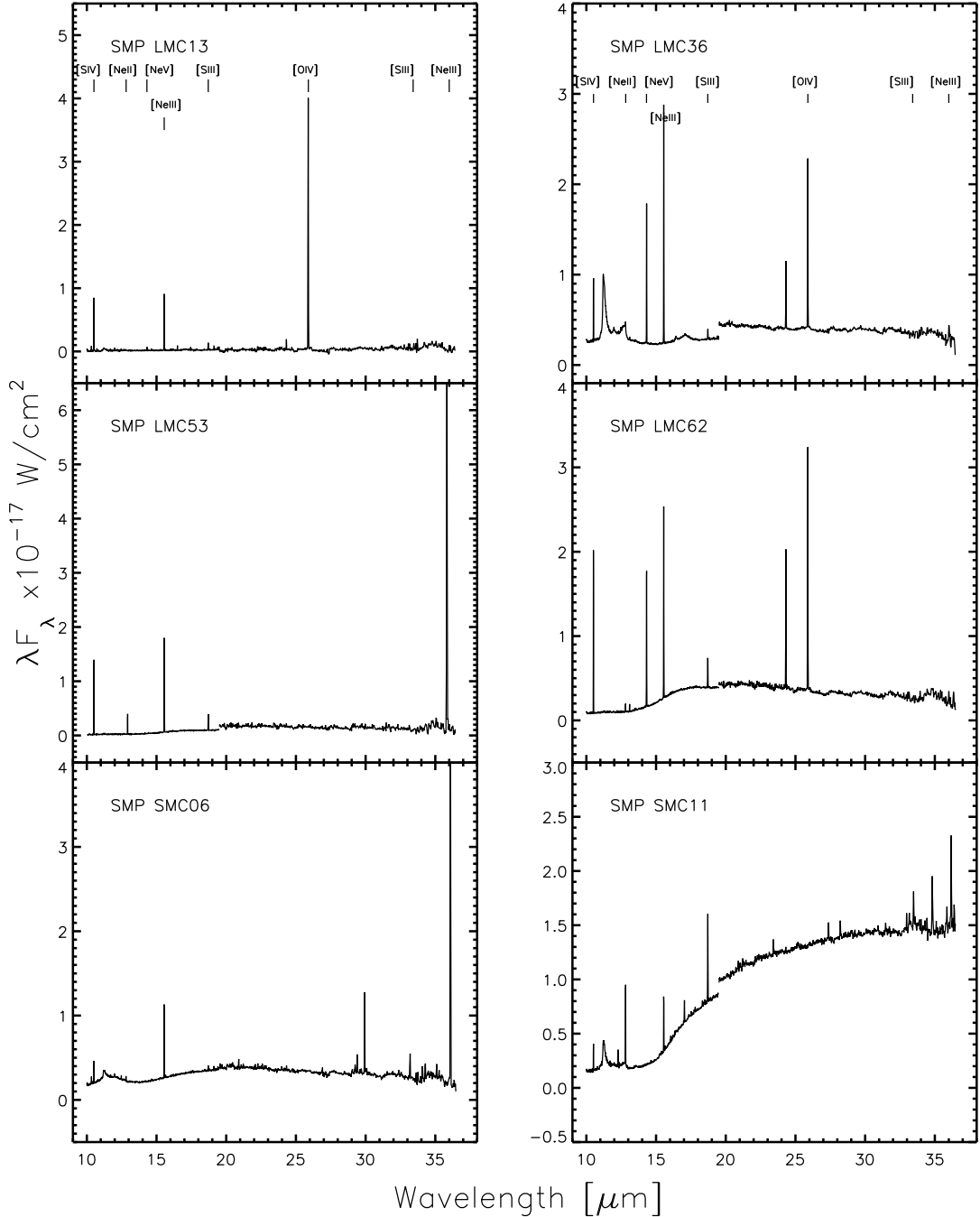}
  \end{center}
  \caption{SH and LH spectra of a handful of PNe. The jump around
    19.5$\mu$m is due to the larger background contribution that falls
    in the LH slit compared to the smaller SH slit (see \S2).\label{spectra_f}}
\end{figure*}

There is an expected mismatch between the SH and LH spectra.  This
mismatch is due to differences in the background contribution that
falls into the slits because the slit size of the LH module is about
4.6 times larger in area than the SH slit. We do not scale the spectra
because we are interested in the line fluxes and the nebulae are
contained in both slits. These PNe are not extended at such a
distances. Diameters of PNe in the LMC given by \citet{sha01} and
\citet{sta99} are usually less than 1\arcsec, and only in a very few
cases does the diameter reach 3\arcsec~(still smaller than the SH slit
width of 4.5\arcsec). Representative examples of the full extracted
high-resolution spectra from 10-37~$\mu$m are shown in
Figure~\ref{spectra_f}. 

\section{Analysis}

  \subsection{Line emission}

  Figure~\ref{lines_f} shows an inset of the most relevant lines for
  the abundance determination present in the spectra of all the
  objects.  The sample ranges from PNe showing high excitation to low
  excitation lines (e.g.  SMP~LMC~85, SMP~LMC~62). Most of the spectra
  show features (PAHs) characteristics of carbon-rich material (e.g.
  SMP~LMC~36, SMP~SMC~11 in Fig.~2), except for SMP~LMC~53 and
  SMP~LMC~62 (Fig.~\ref{spectra_f}) which show amorphous silicates in
  emission at 9 and 18 $\mu$m which is usually an indicator of an
  oxygen-rich environment.

\tabletypesize{\small}

\begin{deluxetable*}{c c c c c c c c}
  \tablewidth{0pc} 
  \tablecaption{Line fluxes\tablenotemark{a}~ of the observed LMC and SMC PNe.\label{line_t}} 

  \tablehead{\colhead{Object} & \colhead{\ion{[S}{4]}} &
    \colhead{\ion{[Ne}{2]}} & \colhead{\ion{[Ar}{5]}} &
    \colhead{\ion{[Ne}{5]}} & \colhead{\ion{[Ne}{3]}} &
    \colhead{\ion{[S}{3]}} & \colhead{\ion{[O}{4]}} \\
    \colhead{} & \colhead{(10.51$\mu$m)} & \colhead{(12.89$\mu$m)} &
    \colhead{(13.10$\mu$m)} & \colhead{(14.31$\mu$m)} &
    \colhead{(15.55$\mu$m)} & \colhead{(18.73$\mu$m)} &
    \colhead{(25.89$\mu$m)}}

    \startdata

    SMP LMC~02  &   $<$0.72  &      1.84  &   $<$0.32  &   $<$0.26  &   $<$0.66  &      0.82  &   $<$0.75  \\ 
    SMP LMC~08  &      2.95  &      7.48  &   $<$0.38  &   $<$0.38  &     34.19  &      1.98  &   $<$0.63  \\ 
    SMP LMC~11  &   $<$1.08  &      1.89  &   $<$0.77  &   $<$0.69  &      0.67  &   $<$0.68  &   $<$1.38  \\ 
    SMP LMC~13  &     12.08  &      0.22\tablenotemark{b} &   $<$0.19  &      0.94  &     12.74  &      2.10  &     55.54  \\ 
    SMP LMC~28  &      0.44  &      1.96  &   $<$0.30  &      0.32  &      3.55  &      0.51  &   $<$1.15  \\ 
    SMP LMC~31  &   $<$0.75  &     22.45  &   $<$0.31  &   $<$0.17  &      0.38\tablenotemark{b} &      0.99  &   $<$1.09  \\ 
    SMP LMC~35  &     10.29  &      0.48  &   $<$0.19  &      0.20\tablenotemark{b} &     16.53  &      1.87  &     27.22  \\ 
    SMP LMC~36  &     12.75\tablenotemark{b} &      1.34  &      0.52  &     25.37  &     36.35  &      1.83  &     22.31\tablenotemark{b} \\ 
    SMP LMC~38  &     12.96\tablenotemark{b} &      2.57  &   $<$0.24  &   $<$0.31  &     48.67  &      3.80  &   $<$1.53  \\ 
    SMP LMC~40  &      2.96  &      0.87  &   $<$0.24  &      2.59  &      6.29  &      1.33  &     24.04  \\ 
    SMP LMC~53  &     20.55  &      0.90  &   $<$0.23  &   $<$0.26  &     26.17  &      4.57  &   $<$1.49  \\ 
    SMP LMC~58  &      2.92  &      2.06  &   $<$0.24  &   $<$0.38  &     20.06  &      1.10  &   $<$2.16  \\ 
    SMP LMC~61  &      7.59  &      6.11  &   $<$0.46  &   $<$0.35  &     30.09  &      6.36  &   $<$0.59  \\ 
    SMP LMC~62  &     27.80  &      1.41  &      1.24  &     28.80  &     32.06  &      6.47  &     38.91  \\ 
    SMP LMC~76  &      4.05  &      2.28  &   $<$0.29  &   $<$0.23  &     20.66  &      2.33  &   $<$0.42  \\ 
    SMP LMC~78  &     25.63  &      1.92  &      0.94  &     26.83  &     46.13  &      5.83  &     47.39  \\ 
    SMP LMC~85  &      1.85  &     13.15  &   $<$0.27  &   $<$0.40  &     16.53  &      2.33  &   $<$0.86  \\ 
    SMP LMC~87  &     11.39  &      3.63  &      0.75  &     12.53  &     11.28  &      4.58  &     39.53  \\ 
                &            &            &            &            &            &            &            \\ 
    SMP SMC~01  &   $<$0.53  &      8.10  &   $<$0.29  &   $<$0.34  &      2.89  &      0.63  &   $<$0.66  \\ 
    SMP SMC~03  &      2.11  &   $<$0.35  &   $<$0.32  &   $<$0.24  &      3.09  &   $<$0.48  &   $<$1.25  \\ 
    SMP SMC~06  &      4.41  &      0.95  &   $<$0.23  &   $<$0.36  &     14.87  &      0.92  &   $<$0.65  \\ 
    SMP SMC~11  &      3.76  &     13.44  &   $<$0.30  &   $<$0.37  &      8.19  &     12.67  &   $<$1.18  \\ 
    SMP SMC~22  &      1.67  &      1.30  &   $<$0.30  &      3.60  &      2.53  &      1.25\tablenotemark{b} &      7.44  \\ 
    SMP SMC~24  &      1.52  &      1.82  &   $<$0.33  &   $<$0.25  &      7.46  &      2.28  &   $<$0.84  \\ 
    SMP SMC~28  &      1.53\tablenotemark{b} &      0.43  &   $<$0.24  &      2.53  &      1.70  &      0.47  &      1.29\tablenotemark{b} \\ 

     \enddata
    
     \tablenotetext{a}{Fluxes in units of $\times$10$^{-14}$
       erg~cm$^{-2}$~s$^{-1}$. Unless otherwise indicated, the
       uncertainties are less than 10\% for all the lines except the
       \ion{[S}{4]} line flux which has an uncertainty between 10 and
       20\%.}

     \tablenotetext{b}{These lines have uncertainties in the flux
       between 20 and 30\%, except for the \ion{[O}{4]} line in
       SMP~SMC~28 where the error is 42\%.}
  \end{deluxetable*}

\tabletypesize{\normalsize}

\begin{figure*}
  \begin{center}
    \includegraphics[width=13cm]{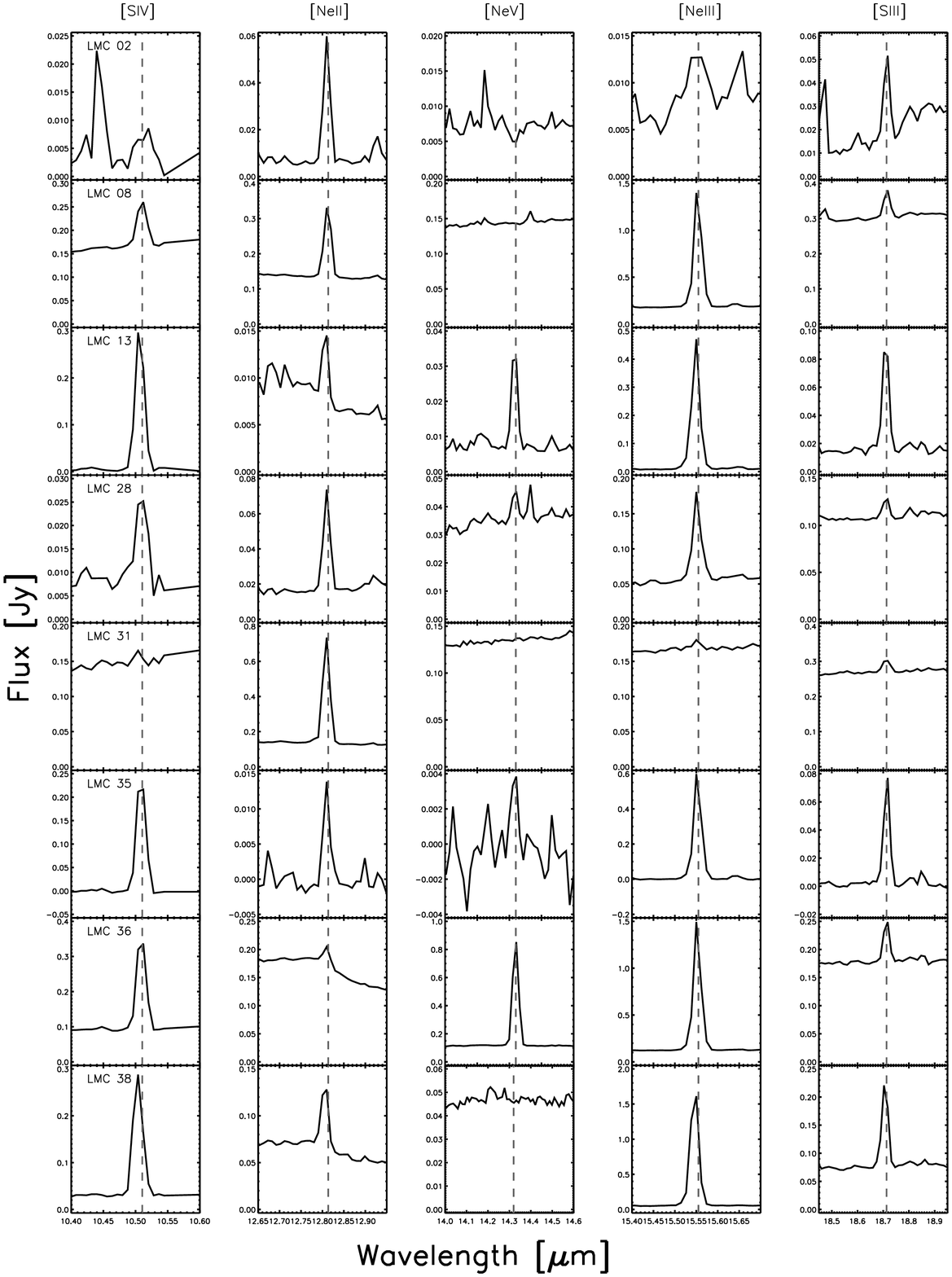}
  \end{center}
  \caption{Inset of the \ion{[S}{4]}, \ion{[Ne}{2]}, \ion{[Ne}{5]},
    \ion{[Ne}{3]}, and \ion{[S}{3]} fine-structure lines for each
    object, except for SMP~LMC~11 because this spectrum was shown in
    \citet{ber06}. The vertical dashed lines indicate the nominal
    position of the lines in the vacuum.\label{lines_f}}
\end{figure*}

\setcounter{figure}{2}

\begin{figure*}
  \begin{center}
    \includegraphics[width=13cm]{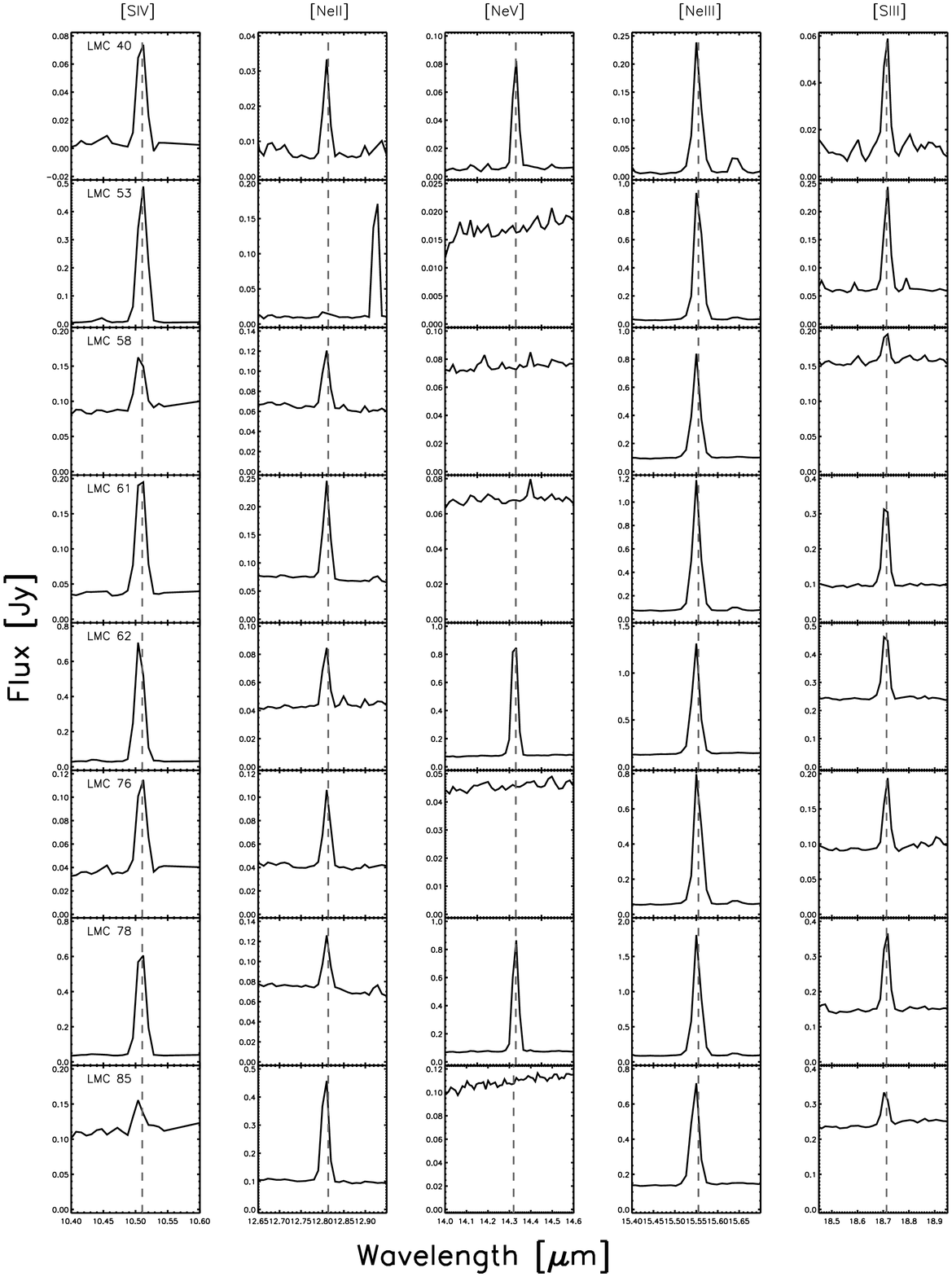}
  \end{center}
  \caption{Continued.}
\end{figure*}

\setcounter{figure}{2}

\begin{figure*}
  \begin{center}
    \includegraphics[width=13cm]{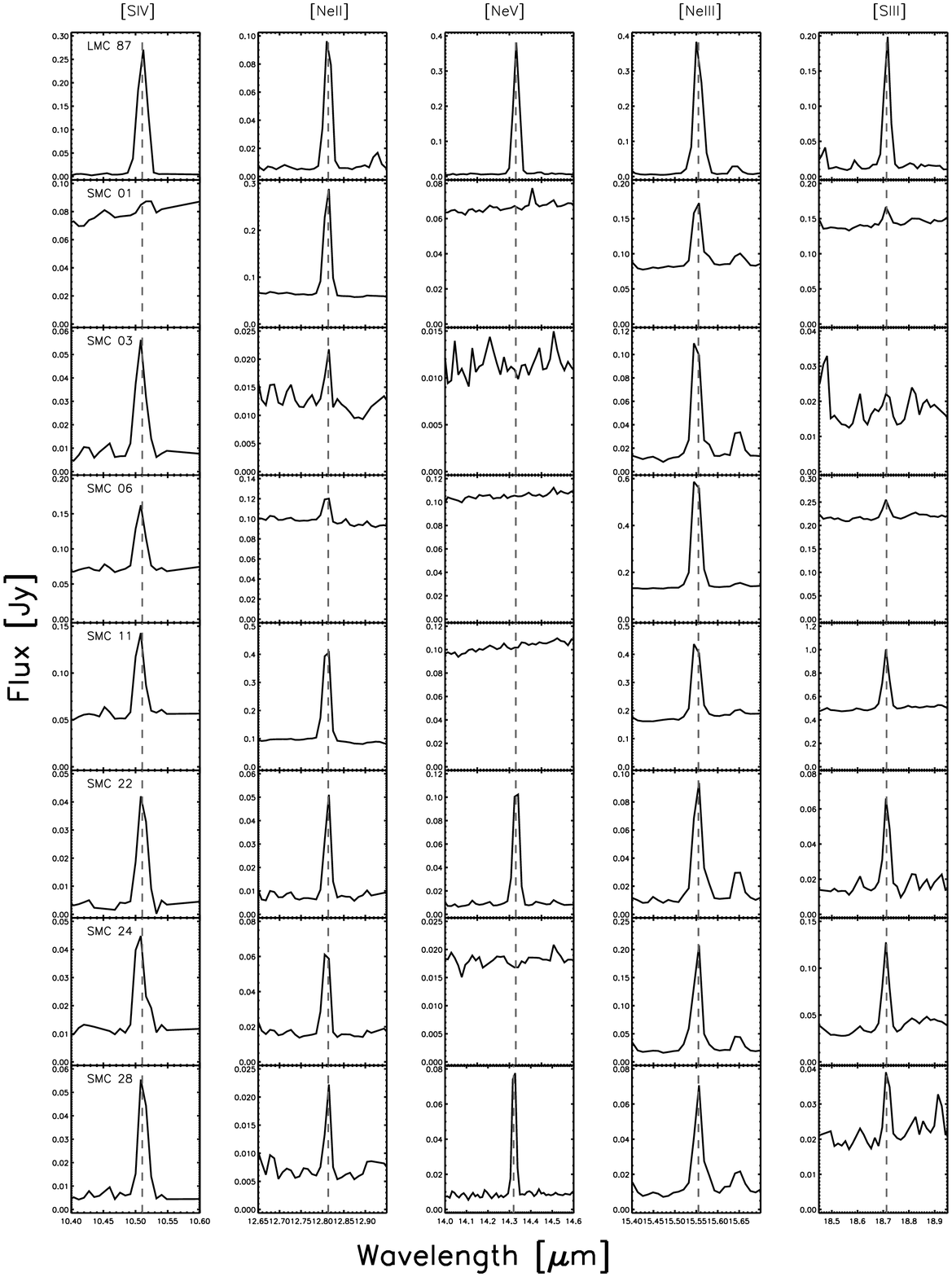}
  \end{center}
  \caption{Continued.}
\end{figure*}

%

The line fluxes are listed in Table~\ref{line_t}. In addition to the
lines listed in this table, other lines such as \ion{[Mg}{5]} at
13.52~$\mu$m have been measured for the PNe with higher S/N. The
\ion{[S}{3]} 33.48~$\mu$m line is always detected when the 18.7~$\mu$m
is present, but because the spectrum is noisier at the long wavelength
end of the LH module (see Fig.\ref{spectra_f}) we preferred to use the
18.71~$\mu$m line flux for the abundance determination of this ion
which also has a larger transition probability and critical density.
Similarly we favored the use of the \ion{[Ne}{3]} 15.55~$\mu$m line
instead of the 36.0~$\mu$m line when both were measured.
Table~\ref{line_t} also includes (mostly) upper-limits on the argon
line at 13.1~$\mu$m.  Although not shown in this paper, the
\ion{[Ar}{3]} line at 8.99~$\mu$m is detected in 9 objects in the
low-resolution spectra, and the \ion{[Ar}{2]} 6.99~$\mu$m line in 2
objects.  The line fluxes were measured using the Gaussian
line-fitting routine in {\it Smart}.  These were derived for each nod
independently. The uncertainty in the fluxes was assumed to be the
largest of either the difference between the flux in the {\em nod}
positions, or the uncertainty in the fit.  These errors are given as
footnotes in the Table. The upper-limits were calculated from a
Gaussian fit with height 3 times the {\em rms} and a FWHM as given by
the resolution of the instrument.

  \subsection{Assumed parameters}

  Table~\ref{tene_t} lists the H$\beta$ flux, extinction, electron
  density (N$_{e}$) and electron temperature (T$_{e}$) assumed to
  calculate the abundances. These values were compiled from a large
  list of references which are given in the footnote of the table.
  When several values were given by different authors an average was
  used. Only in the cases where several values differed by a large
  amount (i.e. SMP~LMC~87, SMP~SMC~11), we assumed those we estimate
  are more accurate. The infrared lines are little affected by
  uncertainties in the T$_{e}$ and extinction corrections but the
  abundances relative to hydrogen are derived using optical
  measurements of H$\beta$ which are affected by these factors. The
  extinction law used is that of \citet{flu}. In two cases no N$_{e}$
  was reported and we assumed a value of 3000~cm$^{-3}$ which seems a
  reasonable value in view of the other measurements.

  \subsection{Abundance determination}

  Using the above parameters, the ionic abundances were computed from
  the infrared line intensities using Eq.~1 of \citet{ber01}. The
  results are shown in Table~3. The total abundances are also given in
  this table, where sometimes a correction due to missing ionization
  stages is necessary.

  For the sulfur abundance the addition of S$^+$, S$^{+2}$ and
  S$^{+3}$ is sufficient for those nebulae for which no \ion{O}{4} is
  observed. The S$^{+}$ abundance was determined using the optical
  lines of S$^{+}$ measured by \citet{mea91a, mea91b} and
  \citet{sta02,sta03}.  The contribution from this ion is usually
  small and in the order of what it is found in Galactic PNe. For the PNe which show the
  \ion{O}{4} line we must take into account the possibility that
  S$^{+4}$ is present.  An estimate may be obtained by looking at
  the two photo-ionization models of Me2-1 \citep{sur04} and NGC 6886
  \citep{pot05}. Both of these PNe are excited by high temperature
  stars (with T$_{eff}$ between 140\,000 and 180\,000~K) and both show
  \ion{O}{4} lines. In addition, Me2-1 has lower average abundances
  compared to Galactic PNe which are closer to the nebulae studied
  here.  Both models give similar results and show that S$^{+4}$
  contributes between 7 and 23\% of the total sulfur abundance based
  on the S$^{+4}$/S$^{+3}$ ratio.  We have used these numbers to
  correct the sulfur abundances in Table~3.

  In the case of neon no \ion{Ne}{4} line has been observed. For those PNe
  which do not show an \ion{O}{4} line it is unlikely that there is
  any Ne$^{+3}$ because it requires a higher energy radiation field
  than does \ion{O}{4}. For those PNe which do show the \ion{O}{4}
  line (7 in the LMC and 2 in the SMC) a correction must be made. This
  can be done in two ways. First, using the same models as for sulfur
  we obtain a contribution of Ne$^{+3}$ that varies from 2 to 33\%
  of the total neon abundance depending on the strength of the
  \ion{Ne}{5} line. A second way of determining the correction could
  be done by looking at the neon abundances in the sample studied by
  \cite{pot06}. This study made use of the same infrared lines as used
  here and had both \ion{Ne}{3} and \ion{Ne}{5} lines, but the
  ultraviolet lines of \ion{Ne}{4} at 2422 \AA~were also observed so
  that Ne$^{+3}$ could also be determined. For these PNe the
  Ne$^{+3}$ abundance was on average 21\% of the total neon
  abundance and 35\% of the sum of Ne$^{+2}$ and Ne$^{+4}$.  To be
  consistent with the sulfur abundance we have used the first method,
  but the results using the second method would result in the same
  values within the expected errors. The total neon abundances are
  given in Table~3.

  The uncertainty in the values of both the Ne$^{+3}$ and S$^{+4}$
  abundances is probably not more than a factor of 2. This leads to a
  maximum error for sulfur and neon of about 30\% from this source. In
  addition uncertainties of measurement of the other ions in the
  infrared are about 10\% with only a few exceptions (see footnote in
  Table~2). The error in the optical lines we used is of the same
  order. In addition an error in the extinction affects the H$\beta$
  flux and uncertainties in the electron temperature dominate the
  uncertainty in our abundance determination. By comparing the optical
  measurements for the same objects by different observers in the
  literature, which usually agree, we estimate that the total error
  remains within 50\% except for SMP~LMC~08 and SMP~SMC~11. The
  abundances of SMP~LMC~08 are uncertain probably because of the assumed
  H$\beta$ flux. (see \S5.3). In SMP~SMC~11 the uncertainty is dominated by
  the inconsistent H$\beta$ flux and extinction quoted in the literature
  and for the purpose of this paper the most recent values given in
  the literature have been adopted.

\begin{deluxetable*}{c c c c c c c c c c c c}
\tabletypesize{\small}
  \tablecolumns{12}
  \tablewidth{0pc} 
  \tablecaption{Neon and Sulfur abundances\tablenotemark{a}.\label{abun_t}} 

  \tablehead{\colhead{Object} & \colhead{Ne$^{+}$/H} &
    \colhead{Ne$^{+2}$/H}  & \colhead{Ne$^{+4}$/H} &
    \colhead{ICF} & \colhead{Ne/H} & \colhead{S$^{+}$/H} &
    \colhead{S$^{+2}$/H} & \colhead{S$^{+3}$/H} &
    \colhead{ICF} & \colhead{S/H} & \colhead{Ne/S}}

  \startdata

SMP~LMC~02  &    3.14 &$<$0.55 &        &   1    &     3.41 &          &   1.26 &$<$0.22 &   1     &   1.37 &   25 \\ 
SMP~LMC~08\tablenotemark{c}  &   32.10 &  73.60 &        &   1    &   105.70 &          &   9.06 &   2.64 &   1     &  11.70 &   90 \\ 
SMP~LMC~11  &    4.56 &        &        &   1    &     4.56 &          &        &        &         &        &       \\
SMP~LMC~13  &    0.17 &   4.81 &   0.05 &   1.02 &     5.12 &     0.14 &   1.46 &   1.75 &   1.20  &   4.01 &   13 \\ 
SMP~LMC~28  &    4.68 &   4.11 &   0.45 &   1.09 &    10.06 &     0.79 &   1.06 &   0.19 &   1     &   2.03 &   49 \\ 
SMP~LMC~31  &    6.95 &   0.06 &        &   1    &     7.01 &     0.13 &   0.34 &$<$0.05 &   1     &   0.47 &  148 \\ 
SMP~LMC~35  &    0.35 &   5.86 &        &   1    &     6.21 &     0.11 &   1.10 &   1.28 &   1.19  &   2.97 &   21 \\ 
SMP~LMC~36  &    0.32 &   4.36 &   0.46 &   1.18 &     5.96 &          &   0.38 &   0.58 &   1.23  &   1.18 &   51 \\ 
SMP~LMC~38  &    0.84 &   8.25 &        &   1    &     9.09 &     0.29 &   1.62 &   1.04 &   1     &   2.95 &   31 \\ 
SMP~LMC~40  &    1.19 &   4.19 &   0.24 &   1.08 &     6.06 &     0.53 &   1.45 &   0.68 &   1.10  &   2.92 &   21 \\ 
SMP~LMC~53  &    0.34 &   4.95 &        &   1    &     5.29 &     0.29 &   1.61 &   1.51 &   1     &   3.41 &   16 \\ 
SMP~LMC~58  &    0.74 &   3.95 &        &   1    &     4.69 &     0.05 &   0.79 &   0.35 &   1     &   1.18 &   40 \\ 
SMP~LMC~61  &    1.54 &   4.34 &        &   1    &     5.88 &     0.79 &   3.99 &   0.75 &   1     &   5.53 &   11 \\ 
SMP~LMC~62  &    0.27 &   3.11 &   0.44 &   1.23 &     4.61 &     0.22 &   1.15 &   1.22 &   1.18  &   3.05 &   15 \\ 
SMP~LMC~76  &    0.48 &   2.33 &        &   1    &     2.81 &     0.19 &   0.79 &   0.24 &   1     &   1.22 &   23 \\ 
SMP~LMC~78  &    0.57 &   6.87 &   0.60 &   1.14 &     9.11 &     0.17 &   1.63 &   1.50 &   1.17  &   3.87 &   24 \\ 
SMP~LMC~85  &    2.70 &   1.98 &        &   1    &     4.68 &     0.40 &   1.37 &   0.17 &   1     &   1.94 &   24 \\ 
SMP~LMC~87  &    1.59 &   2.45 &   0.46 &   1.20 &     5.32 &     0.49 &   1.55 &   0.90 &   1.12  &   3.28 &   16 \\ 
$<$LMC$>$\tablenotemark{b} &          &        &        &        &     6.03 &          &        &        &        &     2.72 &     2.2  \\
SMP~SMC~01  &    3.97 &   0.73 &        &   1    &     4.70 &     0.04 &   0.41 &$<$0.06 &   1     &   0.46 &  103 \\ 
SMP~SMC~03  & $<$0.58 &   2.58 &        &   1    &     2.87 &     0.18 &$<$0.35 &   0.71 &   1     &   1.07 &   29 \\ 
SMP~SMC~06  &    0.30 &   2.48 &        &   1    &     2.78 &     0.07 &   0.44 &   0.39 &   1     &   0.90 &   31 \\ 
SMP~SMC~11\tablenotemark{c}  &    1.62 &   0.48 &        &   1    &     2.10 &     0.01 &   1.16 &   0.08 &   1     &   1.25 &   17 \\ 
SMP~SMC~22  &    0.76 &   0.74 &   0.17 &   1.21 &     1.98 &     0.23 &   0.58 &   0.18 &   1.07  &   1.06 &   19 \\ 
SMP~SMC~24  &    0.94 &   1.87 &        &   1    &     2.81 &     0.08 &   0.95 &   0.13 &   1     &   1.17 &   24 \\ 
SMP~SMC~28  &    0.37 &   0.76 &   0.21 &   1.33 &     1.71 &     0.23 &   0.45 &   0.31 &   1.12  &   1.10 &   16 \\ 
$<$SMC$>$ &          &        &        &        &     2.71 &          &        &        &         &   1.00 &   27 \\

  \enddata
  \tablenotetext{a}{The neon and sulfur abundances are in
    $\times$10$^{-5}$ and $\times$10$^{-6}$ respectively.}
  \tablenotetext{b}{The average also includes SMP~LMC~83 from \citet{ber04}.}
  \tablenotetext{c}{Large uncertainty in the abundance (see \S3.3).}
\end{deluxetable*}

\section{Comparison Sources}

This section describes the sources to which the MC PN abundances are
compared in \S5.2. These comparison sources include Galactic PNe,
Galactic, MC, M33 and M83 H\,II regions, and the solar abundance.

  \subsection{Solar values}

  The solar carbon, oxygen, sulfur, argon and neon abundances have
  been subject to significant changes, especially during the last
  seven years. These changes reflect in some way the difficulty in
  deriving solar abundances. The neon and argon abundances are
  especially troublesome because there are no lines of these elements
  in the solar photosphere and their abundances must be derived from
  coronal lines.  \citet{asp05} quoted a neon value of
  6.9$\times$10$^{-5}$ using the oxygen solar abundance and assuming a
  ratio of the Ne/O of 0.15.  Previously, \citet{fel} using coronal
  line measurements found a neon abundance of 1.2$\times$10$^{-4}$.
  This value is much higher than the value by \citet{asp05} but agrees
  better with the earlier values reported by \citet{gre98}.  This
  discrepancy is important in the debate over the consistency of the
  helioseismological measurements and the solar model \citep{ant,bah}.
  The neon abundance derived by \citet{pot06} in a sample of Galactic
  PNe is more consistent with the higher neon value of \citet{fel}.
  Very recently \cite{lan07} derived a value of 1.29$\times$10$^{-4}$,
  again higher than the value given by \citet{asp05} and in very good
  agreement to the previous values reported \cite{fel} and
  \cite{gre98}.  The quoted value of the solar sulfur abundance has
  been decreasing in the last years.  The sulfur abundance derived by
  \citet{gre} is 1.4$\times$10$^{-5}$ while \citet{asp05} find
  0.94$\times$10$^{-5}$.  Given these discrepancies, in the rest of
  the paper instead of assuming a certain value we will refer and
  compare to the above range of solar values.

  \subsection{PNe and H\,II regions}
  
  For comparison purposes we have selected a sample of PNe and H\,II
  regions for which abundances were also derived from infrared data
  and in a similar way to the PNe presented in this paper. The
  Galactic PNe abundances in \citet{pot06} using ISO data have been
  complemented with the Spitzer derived abundances of IC\,2448
  \citep{shannon}, M\,1-42 \citep{pot07}, and NGC\,2392 (Pottasch et
  al., in prep).  Galactic and MC H\,II regions were taken from
  \citet{leti} and \citet{ver} respectively. They include ISO derived
  abundances from 26 H\,II regions in the Milky Way, 13 in the LMC,
  and 3 in the SMC. The Spitzer abundances in Lebouteiller et al.\,(in
  prep) of the giant H\,II regions NGC\,3603 (in the MW), 30 Doradus
  (LMC), and NGC\,346 (SMC) are also included. While Lebouteiller et
  al. (in prep) derive abundances at several positions in each region,
  the calculated abundances are very similar for a given region and
  here we adopt their average value.  \citet{rub07} derived recently
  the Ne/S abundance ratio of H\,II regions in M83 using Spitzer data
  but the absolute values are not given. The same authors are working
  on a study of H\,II regions in M33 and we use their neon and sulfur
  abundance ranges\footnote{These abundances were presented in the
    Xiang Shan workshop in 2007 (http://ast.pku.edu.cn/~xs2007/).} in
  Figure~5.

\section{Discussion}

  \subsection{Neon, Sulfur and the Ne/S ratio}

  Neon and sulfur are alpha-process elements and therefore 
  should track each other.  Their abundances are representative of the
  chemical composition of the cloud from which they formed. Recent
  work has suggested that some PNe may produce neon in the course of
  evolution \citep{mar}. \cite{lei06} claim that because their derived
  oxygen and neon abundances correlate with each other that when oxygen
  is enriched neon must be also self-enriched in their MC sample.
  According to the theoretical models of \citet{kar03} this enrichment
  is modest and the mass range at which neon production in low- and
  intermediate mass stars may occur is very small (around 3\msun).
  Thus statistically few PNe should experience such an effect.

  From the LMC abundances listed in Table~\ref{abun_t} SMP~LMC~08
  shows unusually high abundances of neon and sulfur.  Except for
  SMP~LMC~11, this PN has the lowest H$\beta$ flux of the LMC sample
  (see Table~\ref{tene_t}), and it is likely that the high abundances
  are in part the result of this low H$\beta$ flux which we use to
  derive the abundance. The neon enrichment predicted by \citep{kar03,
    mar} is not large enough to predict such high values.  The lowest
  neon abundance corresponds to SMP~LMC~76 which also has a low sulfur
  abundance compared to the rest of objects (but not the lowest). The
  neon abundance of SMP~LMC~11 has been derived using only the
  Ne$^{+}$ stage of ionization. This is a very low excitation
  object, and has been described as a pre-planetary nebula by
  \cite{ber06} due to its very peculiar infrared spectrum which shows,
  among others features, molecular absorption bands of acetylene and
  poly-acetylene.  There are no \ion{[Ne}{3]} or \ion{[S}{3]} and
  \ion{[S}{4]} lines in the IRS spectrum.

  Not including SMP~LMC~08, the average neon abundance of the PNe in
  the LMC is 6.0$\times$10$^{-5}$, which is 1/2.7 of the average neon
  abundance of Galactic PNe (1.6$\times$10$^{-4}$) used as comparison
  (\S4.2).  The mean sulfur abundance is 2.7$\times$10$^{-6}$, which
  yields a slightly lower ratio of 1/3.7 when compared to the average
  value of Galactic PNe (0.99$\times$10$^{-5}$).  The SMC sample
  contains only 7 objects. Although this is a small number to attempt
  statistics it can be seen in Table~3 that the sulfur and especially
  the neon abundances are very similar in most of the SMC objects.
  The mean neon abundance of the SMC PNe is 2.7$\times$10$^{-5}$,
  which is 1/6.0 of the average neon abundance in Galactic PNe. The
  sulfur abundance in the SMC PNe is also low, 1/10 of the average
  Galactic PNe sulfur abundance. Thus, keeping in mind that the SMC
  sample contains only a few objects, it seems that either the sulfur
  abundance in MC PNe is lower than in Galactic PNe or the neon
  abundance higher.

\begin{figure}
    \begin{center}
      \includegraphics[width=9cm]{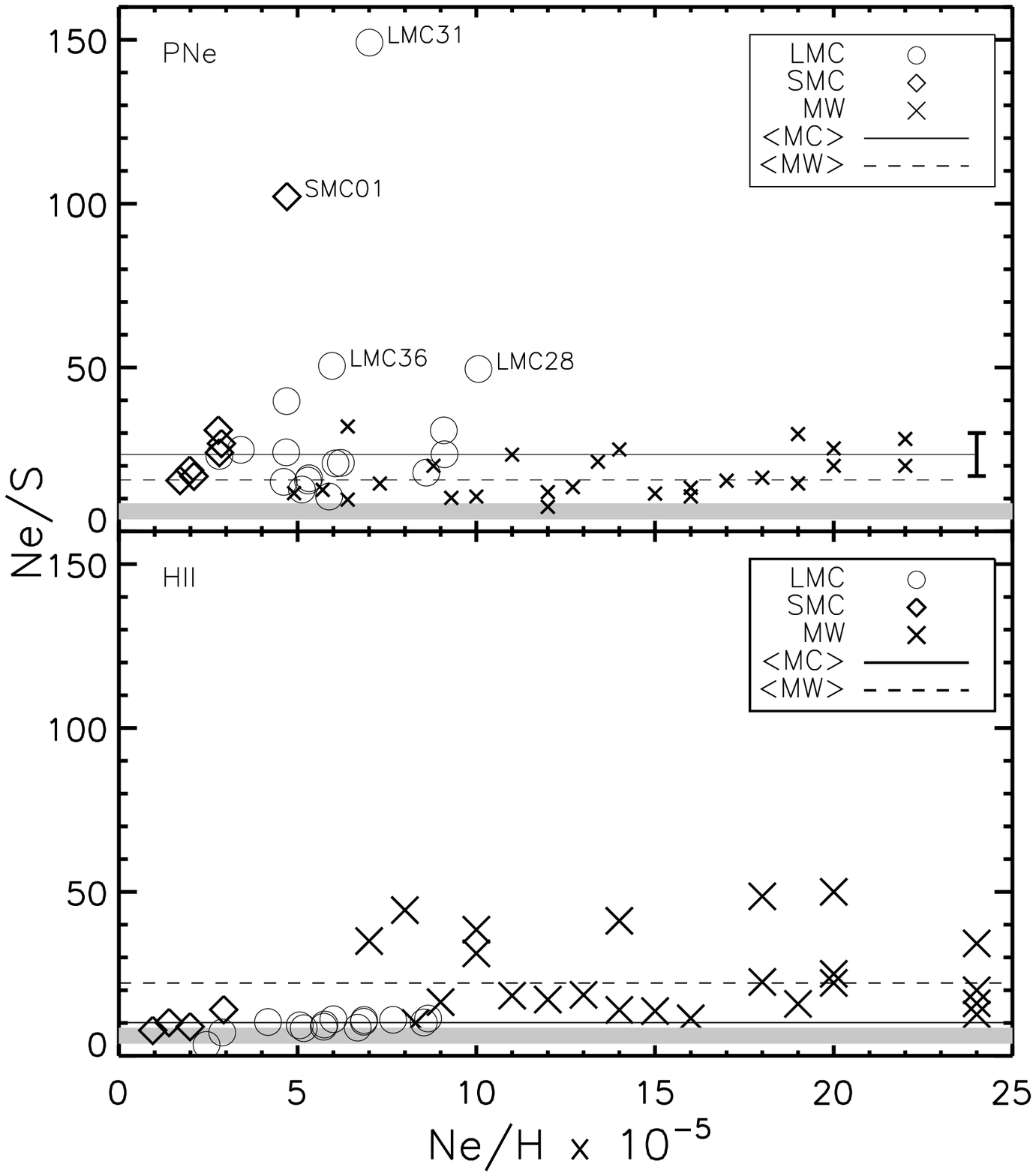}
    \end{center}
    \caption{Ne/S ratio with respect to the total neon abundance of
      the PNe (top panel) and H\,II regions (bottom). The grey band
      represents the solar Ne/S ratio. The horizontal lines represent
      the median of the Ne/S for PNe (with a typical error bar on the
      right side of the line) and H\,II regions in the Galaxy, the
      LMC, and the SMC.\label{nes_f}}
  \end{figure}

  Figure~\ref{nes_f} is a plot of the Ne/S ratio against the neon
  abundance (as an indicator of metallicity) for PNe (top) and H\,II
  regions (bottom).  This abundance ratio has the advantage over the
  elemental abundance measurements in that any uncertainty introduced
  by combining infrared measurements with optical hydrogen measurements
  cancels out, and uncertainties in T$_{e}$ are much reduced. Also
  plotted are the median values for the different data sets. The
  figure shows that most of the MC PNe have ratios between 15 and 30
  (with a median value of 23.5), with SMC and LMC PNe displaying a
  similar range of values. This range is similar to the ratio
  displayed by Galactic PNe (7.5-32), although as we saw before the MC
  PNe have a slightly higher ratio (solid and dashed lines in the
  figure). The Ne/S ratio in Galactic H\,II regions ranges from 10 to
  50 with a median value of 21 (dotted-dashed line in
  Fig.~\ref{nes_f}) and agrees very well with the MC PNe. The MC H\,II
  regions have on average a lower Ne/S ratio but the number of sources
  is not very large.  The H\,II regions studied by \citet{rub07} in
  the metal-rich galaxy M83 have Ne/S ratios that vary from 24.4 to
  41.9 and thus are similar to the ratios shown by the MC PNe. In
  summary, the Ne/S ratios show large variations within each dataset.
  The fact that most LMC/SMC PNe show values which are similar to
  Galactic PNe and different H\,II regions implies that these elements
  have a common origin and if any enrichment of neon has occurred it
  has remained modest. This agrees with the result of \cite{dop97} who
  find no sign of dredge-up of $^{22}$Ne in their sample of LMC PNe.

  There are two PNe with a Ne/S ratio which is significantly higher
  than any of the comparison datasets. Two other objects have a ratio
  of $\sim$50 which is high compared to the rest of the sample of PNe
  although some Galactic H\,II regions also reach such values.  These
  objects are labeled in Figure~4 (top).  The extremely high values of
  the Ne/S ratio in SMP~LMC~31 and SMP~SMC~01 are mainly the result of
  their lower sulfur abundance.  These two objects show the lowest
  sulfur abundances: SMP~LMC~31 has a sulfur abundance which is 5.7
  times lower than the average LMC PNe, and the sulfur abundance in
  SMP~SMC~01 is about 2.2 lower than the average SMC PNe (see
  Table~3).  These differences in sulfur can account for the high
  ratio observed in this two objects. However, these two objects
  together with SMP~LMC~28 also show a high neon abundance compared to
  the rest of objects in their sample and it is therefore possible
  that neon-enrichment has taken place in these objects.
  
  The grey band in Figure~\ref{nes_f} indicates the lower and higher
  solar ratio found using the abundances given in \S3.1 and it is
  clearly lower than Galactic and MC PNe or H\,II regions. This is
  known and several authors have already discussed that the solar
  sulfur abundance seems too high compared to Galactic PNe and H\,II
  regions \citep{pot06,mar, leti}. In addition, several authors
  \citep{pot06,wan07} favor the higher neon solar abundance given by
  \citet{fel} and \citet{lan07} instead of the one quoted by
  \citet{asp05}. This work also supports the higher neon abundance in
  the literature but from Figure~4 it is clear that the Ne/S in most
  datasets is higher than the solar ratio.
 
  \subsection{Comparison to other sources}

  \begin{figure}
    \begin{center}
      \includegraphics[width=7cm,angle=90]{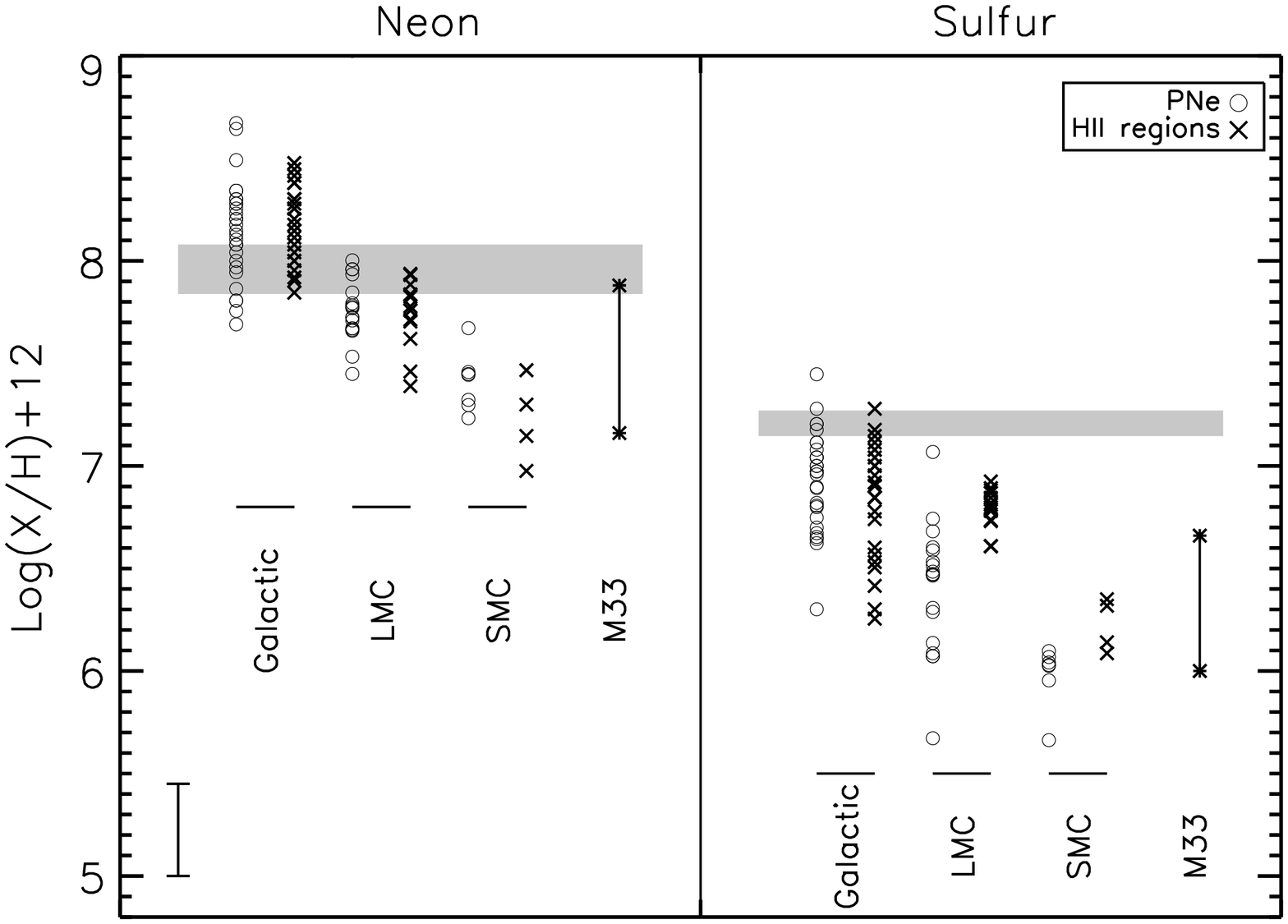}
    \end{center}
    \caption{Comparison of the neon and sulfur abundances in PNe and
      H\,II regions for the Milky Way, the LMC and the SMC. The range
      of abundances for H\,II regions in M33 are also shown. A 50\%
      error bar for the LMC and SMC PNe is shown in the lower side of
      the figure (see \S3.3).\label{abun_f}}
  \end{figure}

  Figure~\ref{abun_f} shows a comparison of the neon and sulfur
  abundances in PNe and H\,II regions in the MW, LMC, SMC and M33.  The
  grey band in the figure represents the range of solar values found in
  the literature.

  Galactic PNe and H\,II regions display a neon abundance that is
  closer to the higher solar value. Note that while we are comparing
  with solar values the effect of the Galactic abundance gradient has
  not been taken into account in this plot. The Galactic sources also
  show a clear under-abundance of sulfur compared to the solar value.
  The neon abundances in PNe and H\,II regions show a remarkable
  agreement in the Milky Way, LMC and SMC.  M33 which is usually
  regarded as having half the solar metallicity \citep{mag07} shows a
  very large range of neon abundance which encompasses both the LMC
  and SMC PNe sample.

  The interpretation of the sulfur abundance is complicated. As
  mentioned before, both Galactic PNe and H\,II regions show clearly
  an under-abundance of sulfur with respect to solar, and this has
  been ascribed in some studies to the solar sulfur abundance being
  too high.  The LMC PNe sulfur abundance shows a very large
  dispersion in values. Most PNe show sulfur values that are in the
  range of the LMC H\,II regions but 6 objects are below that range.
  This does not mean that these PNe have anomalous sulfur abundances
  as the number of LMC H\,II regions used for comparison is small, but
  seems to point to a slightly lower sulfur abundance in the PNe.  The
  low sulfur abundances shown by these six PNe are not due to errors
  in the ICF used to correct for S$^{+4}$ because some of them are
  in fact low ionization PNe and no ICF is needed.  We have compared
  the average abundance of sulfur in high- and low-excitation PNe
  separately and find similar average abundances of 2.9 and
  2.3$\times$10$^{-6}$ for the high- and low-excitation PNe
  respectively.  Except for SMP~LMC~31, the range of sulfur abundances
  in the H\,II regions in M33 (which has a metallicity close to that
  of the LMC) is similar to the sulfur abundances we measure in these
  PNe.  The comparison of the PNe and H\,II region sulfur abundances
  in the SMC is hampered by the low number of sources.  Considering
  the uncertainties both data sets compare well, although one source
  (SMP~SMC~01) has a sulfur abundance which is clearly lower than the
  rest of PNe and H\,II regions. We do not know the reasons for the
  apparent sulfur depletion in SMP~LMC~31 and SMP~SMC~01. Sulfur can
  be depleted onto dust (e.g. MgS, FeS).  Although both SMP~LMC~31 and
  SMP~SMC~01 do show a strong MgS feature a further investigation of
  this possibility must be made before any conclusion can be drawn.

  \subsection{Comparison with the literature}

\begin{deluxetable*}{c c c c c c c c c c}
  \tabletypesize{\small}
  \tablecolumns{10}
  \tablewidth{0pc} 
  \tablecaption{Abundance Comparison.\label{comp_t}} 
  \tablehead{\colhead{Object} &
    \multicolumn{4}{c}{Ne/H$\times$10$^{-5}$} & &
    \multicolumn{4}{c}{S/H$\times$10$^{-6}$}\\ \cline{2-5} \cline{7-10}
    (SMP) & \colhead{Present} & \colhead{DM\tablenotemark{a}} & \colhead{D97\tablenotemark{a}} &
    \colhead{LD\tablenotemark{a}} & & Present &
    \colhead{DM\tablenotemark{a}} & \colhead{D97\tablenotemark{a}} &
    \colhead{LD\tablenotemark{a}}}

  \startdata

  LMC02  &     3.41 &        &  2.8 & $<$0.18  &  &       1.4  &     1.6 &  4.0  &   $<$4.2    \\
  LMC08  &   105.7  &    2.2 &  2.1 &    1.82  &  &      11.7  &     6.0 &  4.9  &   1109.7    \\
  LMC11  &     4.6  &        &      &    0.55  &  &            &         &       &      1.15   \\
  LMC13\tablenotemark{d}  &     5.1  &    3.8 &      &    4.0   &  &       4.0  &     3.9 &       &             \\
  LMC28\tablenotemark{d}  &    10.1  &        &      &          &  &       2.0  &         &       &             \\
  LMC31  &     7.0  &        &      &    0.04  &  &       0.47 &         &       &      1.86   \\
  LMC35\tablenotemark{d}  &     6.2  &    4.0 &  3.0 &    3.0   &  &       3.0  &     2.8 &  6.0  &     24.0    \\
  LMC36\tablenotemark{d}  &     6.0  &        &      &    5.3   &  &       1.18 &         &       &             \\
  LMC38  &     9.1  &    4.1 &      &    3.6   &  &       2.9  &    1    &       &     23.4    \\
  LMC40\tablenotemark{d}  &     6.1  &    6.0 &  4.5 &    6.9   &  &       2.9  &     4.6 &  6.2  &     36.3    \\
  LMC53  &     5.3  &        &      &    4.1   &  &       3.4  &         &       &     18.2    \\
  LMC58  &     4.7  &    2.0 &      &    1.23  &  &       1.18 &     3.1 &       &     95.5    \\
  LMC61  &     5.9  &    4.2 &      &    5.8   &  &       5.5  &     7.4 &       &      5.7    \\
  LMC62\tablenotemark{d}  &     4.6  &    3.3 &      &    2.2   &  &       3.1  &     8.0 &       &      4.6    \\
  LMC76  &     2.8  &    2.4 &  2.2 &    1.86  &  &       1.2  &     4.0 &  4.0  &             \\
  LMC78\tablenotemark{d}  &     9.1  &    4.5 &      &    3.6   &  &       3.9  &     5.2 &       &     30.9    \\
  LMC83  &     8.6  &    4.1 &  5.1 &          &  &       4.8  &     6.5 &  2.4  &             \\
  LMC85  &     4.7  &    4.0 &  3.0 &    0.68  &  &       1.9  &     4.0 &  2.0  &      4.7    \\
  LMC87\tablenotemark{d}  &     5.3  &   11.0 &      &    3.1   &  &       3.3  &     6.8 &       &     18.2    \\
  SMC01  &     4.7  &    0.7 &      &    0.26  &  &       0.46 &         &       &             \\
  SMC03  &     2.9  &    1.3 &      &    1.00  &  &       1.01 &         &       &      5.7    \\
  SMC06  &     2.8  &    2.2 &      &    1.38  &  &       0.90 &     6.0 &       &     23.4    \\
  SMC11  &     2.1  &    1.5 &      & $<$1.00  &  &       1.25 &     2.8 &       &   $<$5.9    \\
  SMC22\tablenotemark{d}  &     2.0  &    2.1 &      &    0.56  &  &       1.06 &     3.2 &       &      4.0    \\
  SMC24  &     2.8  &        &      &    1.15  &  &       1.17 &         &       &      0.31   \\
  SMC28\tablenotemark{d}  &     1.7  &    3.0 &      &    0.72  &  &       1.10 &     4.0 &       &      7.4    \\
  
  \enddata
  \tablenotetext{a}{References to abundances: DM
    \citep{dop91a,dop91b}, D97 \citep{dop97}, LD \citep{lei06}.}
  \tablenotetext{d}{High excitation PNe.}
\end{deluxetable*}

  A comparison of the abundances derived in this paper with those
  derived in \citet{dop91a,dop91b,dop97} using photo-ionization models and
  \citet{lei06} using optical data is given in Table~\ref{comp_t}.

  The neon abundances we derive are in agreement with the abundances
  of \citet{dop91a,dop91b}. In most of the cases the agreement is very
  good. For the few objects where the agreement is not that
  satisfactory, the abundances usually compare well within a factor of
  two which is reasonable considering the uncertainties in the
  abundances we derive and those involved in the use of
  photo-ionization models.  \cite{dop97} revised the abundances for 8
  LMC PNe adding {\em HST} spectroscopy to their ground observations.
  Some of their revised abundances agree with their previous
  determinations but in some cases (especially for sulfur) the
  differences amount to a factor of 1.3 to 3.  Our abundances are
  roughly similar to those of \citet{lei06} for most of the
  high-excitation PNe\footnote{Those PNe showing high-excitation lines
    such as \ion{[Ne}{5]}. See footnote in Table~4.} except for
  SMP~LMC~62 and SMP~LMC~83.  \citet{lei06} however state that their
  neon abundance for these two objects was poorly determined. For most
  of the remaining sources (low-excitation PNe) our abundances are
  significantly larger than those by \cite{lei06}.  To determine the
  reasons for this discrepancy, we have calculated the \ion{Ne}{3}
  abundance using their measured optical line at 3869\AA~and assumed
  T$_e$.  The \ion{Ne}{3} fractional abundance we derive using the
  optical line agrees within 30\% with the results from the infrared
  line for the high-excitation PNe and a few of the low-excitation PNe
  (SMP~LMC~53, SMP~LMC~61, SMP~LMC~76). For the rest of the objects
  the difference in the \ion{Ne}{3} ionic abundance is large enough to
  account for the difference in the total neon abundance. A possible
  explanation for this discrepancy is the uncertainty of T$_e$ when
  deriving abundances using optical lines. In some PNe the T$_e$
  derived using the \ion{Ne}{3} is significantly lower than using
  other ions \citep{ber01}. A lower T$_e$ increases the ionic
  abundance, and a difference of a couple of thousand degrees Kelvin
  in the T$_e$ can account for the differences we see.

  The sulfur abundance reported in this paper agrees (although to a
  lesser extent than the neon) with those of \cite{dop91a,dop91b}.
  The abundances are usually within a factor of 3; when the agreement
  is less good our abundances are always lower.  \citet{dop91a,dop91b}
  have information on the S$^{+}$ and S$^{+2}$ ions, and in
  several cases only upper-limits for one or both ions could be
  derived. However, the contribution to the sulfur abundance in PNe
  comes mainly from the S$^{+2}$ and S$^{+3}$, and both these ions
  can be measured in the infrared.  The abundances by \cite{lei06} are much
  higher than either presented in this paper or by
  \cite{dop91a,dop91b}.  The discrepancies are sometimes larger than a
  factor of 15.  Some of the sulfur abundances reported by
  \cite{lei06} are even higher than solar by factors of a few which is
  difficult to interpret.  This large difference cannot be accounted
  for by differences in the S$^{+2}$ fractional abundance using
  infrared or optical lines which we have recalculated. The \ion{S}{3}
  line at 6312\AA~used by \cite{lei06} is blended with a \ion{He}{2}
  line which may add a fraction to the total sulfur abundance. In
  addition, for many cases \cite{lei06} could only measure
  upper-limits to the \ion{S}{3} line and they state that their sulfur
  abundance is very uncertain.  It is possible that the ICFs used to
  derive the total sulfur abundance in \cite{lei06} may overestimate
  the contribution of S$^{+3}$ by a large factor.  Despite the
  uncertainties explained in our abundance determination we consider
  our sulfur abundances to be more accurate than the previous work as
  we have measured the important stages of ionisation and complemented
  with the existing data in the literature.

  SMP~LMC~08 needs special mention. \citet{lei06} quote a much lower
  abundance of neon (1.8$\times$10$^{-5}$) than the one we derive. We
  find that using their measured optical line at 3869\AA, extinction,
  T$_e$, and N$_e$ values, the Ne$^{+2}$ fractional abundance is a
  factor 32 lower than using the infrared line. The infrared line at
  15.5$\mu$m is very bright (Fig.~3) and could be measured easily.
  While we do not know the nature of this discrepancy, a small part of
  this discrepancy can be ascribed to difference in the extinction,
  since \citet{lei06} use a very low value (C$=$0.01) compared to the
  one by \citet{mea91a} which we use, or uncertainties in the T$_e$.
  The IRS spectrum of the object shows that \ion{Ne}{2} is also an
  important contributor to the total neon abundance (about 30\%) and
  this may be underestimated by \citet{lei06}.  SMP~LMC~08 also shows
  a large abundance of sulfur but interestingly in this case
  \citet{lei06} quote a lower limit to its abundance which is a factor
  10 higher than ours (1.17$\times$10$^{-5}$) and of the order of the
  solar value (1.4$\times$10$^{-5}$). The Ne/S abundance we find is
  about 9, which is similar (but a bit on the lower end) to the rest
  of the sample.  This may imply that the H$\beta$ flux we use for
  this object \citep{woo87} to derive the elemental abundance is
  probably not adequate, and will render the abundance derived for
  this object as uncertain.

\section{Summary and Conclusions}

We report the high-resolution Spitzer-IRS observations of a sample of
18 PNe in the LMC and 7 in the SMC. The spectra cover the 10-37~$\mu$m
wavelength range and show the usual fine-structure lines of neon,
sulfur and oxygen typically seen in PNe. Some nebulae also show high
excitation lines of argon and magnesium.

The abundances for neon and sulfur have been derived and compared to
Galactic PNe and H\,II regions, MC H\,II regions and the solar values.
The neon average abundances are 6.0$\times$10$^{-5}$ and
2.7$\times$10$^{-5}$ in the LMC and SMC respectively. This is
$\sim$1/3 and 1/6 of the average neon abundance of the Galactic PNe
used as comparison. Given the uncertainties, these values agree well
with the most quoted values for the LMC and SMC metallicity (1/3 and
1/5 solar respectively).

The average Ne/S ratio of the MC PNe (23.5) is slightly higher than
the average ratio for Galactic PNe (16) but the range of Ne/S values
is similar in both samples. These values are also similar to those
found in H\,II regions. We believe that this is an indication that
neon-enrichment has either not occurred or remained modest in most of
the nebulae. This agrees with the conclusion by \citet{dop97} based on
their derived abundances in a sample of LMC PNe. In fact,
nucleosynthesis models suggest that this process occurs in a very
narrow range of masses and thus statistically few objects would
experience such enrichment. Four objects show a high Ne/S ratio. In
two of them (SMP~LMC~31 and SMP~SMC~01) the high ratio is mainly due
to the very low sulfur abundance of these objects, but together with
SMP~LMC~28 they show the highest neon abundance in the sample so it
is a possibility that these PNe may have experienced some
neon-enrichment.

The range of neon abundances of PNe and H\,II regions is the same in
the Milky Way, the LMC and the SMC. The sulfur abundance of Galactic
PNe and H\,II regions is also similar for both sets of objects.  For
the LMC it seems that some PNe show lower sulfur abundances than the
H\,II regions, but the number of LMC H\,II regions to which we compare
is not very large. The sulfur abundances of M33 H\,II regions have a
similar range of abundances to those of the LMC and SMC PNe.  Given
the low number of objects in the sample of SMC PNe and H\,II regions
we state that it seems that both types of objects show similar sulfur
abundances but clearly more objects of both kinds are needed.

The two nebulae, SMP~LMC~31 and SMP~SMC~01, showing a clear lower
sulfur abundance compared to the rest of the objects also show a MgS
feature in their low-resolution IRS spectra. One could argue that some
of the sulfur is depleted in dust. However other objects which show
the MgS feature do not have a lower sulfur abundance. This should be
further investigated.

The PNe abundances derived are also compared to previous
determinations by \cite{lei06} from optical line measurements and
\citet{dop91a,dop91b,dop97} using photo-ionization models. The
comparison shows that our derived neon abundance agrees very well
with those by \citet{dop91a,dop91b} and to a lesser extent with those
of \cite{lei06}. The sulfur abundances we derive agree well for about
half of the objects with the abundances determined by
\citet{dop91a,dop91b} but for the rest of the PNe they are up to a
factor of 3 lower. The sulfur abundances derived by \cite{lei06} are
much higher than either the ones derived in this paper or the ones by
\citet{dop91a,dop91b}. The advantage of the abundances presented in
this paper over previous work is that we have measured and used the
most important stages of ionization and complemented with existing
data in the literature.

\acknowledgments We thank an anonymous referee whose comments have
improved the paper. JBS would like to thank Duncan Farrah for reading
parts of the manuscript. This work is based on observations made with
the Spitzer Space Telescope, which is operated by the Jet Propulsion
Laboratory, California Institute of Technology under NASA contract
1407. Support for this work was provided by NASA through Contract
Number 1257184 issued by JPL/Caltech.

\end{document}